\documentclass[%
 reprint,
 amsmath,amssymb,
 aps,
superscriptaddress
]{revtex4-2}

\usepackage{array}
\usepackage{booktabs}
\usepackage{tabularx}
\usepackage{graphicx}% Include figure files
\usepackage{dcolumn}% Align table columns on decimal point
\usepackage{bm}% bold math
\usepackage{hyperref}% add hypertext capabilities
\usepackage{xcolor}
\usepackage{multirow}
\usepackage{float}

\newcommand{\mnras}{Mon.\ Not.\ R.\ Astron.\ Soc.}
\newcommand{\jcap}{J.\ Cosmol.\ Astropart.\ Phys.}
\newcommand{\aap}{Astron.\ Astrophys.}
\newcommand{\physrep}{Phys.\ Rep.}
\newcommand{\pasp}{Publ.\ Astron.\ Soc.\ Pac.}

\newcommand{\Lbox}{L_{\text{box}}}

\newcommand{\xHI}{x_{\text{HI}}}
\newcommand{\xHII}{x_{\text{HII}}}
\newcommand{\xHIIavg}{\overline{x_{\text{HII}}}^{\rm{V}}}
\newcommand{\Tcmb}{T_{\text{CMB}}}

\newcommand{\Mh}{M_{\text{h}}}
\newcommand{\Msun}{M_{\odot}}
\newcommand{\Deltacross}{\Delta^2_{\rm{21cm,}\nu}}
\newcommand{\xicross}{\xi_{\rm{21cm,}\nu}}
\newcommand{\rdelta}{\tilde \rho_{\rm{21cm,}\nu}}
\newcommand{\rxi}{\rho_{\rm{21cm,}\nu}}
\newcommand{\Rb}{R_{\rm{b}}}
\newcommand{\Vb}{V_{\rm{b}}}
\newcommand{\BSD}{\Vb\Rb dn/d\Rb}
\newcommand{\fesc}{f_{\rm{esc}}}
\newcommand{\fesczero}{\fesc^{(0)}}
\newcommand{\alphaesc}{\alpha_{\rm{esc}}}
\newcommand{\LX}{L_{\rm{X,SFR}}}

\newcommand{\olimpus}{\texttt{oLIMpus}}
\newcommand{\zeus}{\texttt{Zeus21}}

\begin{document}

\title{Reionization Bubbles from Real-Space Cross Correlations of Line Intensity Maps}

\author{Emilie Thélie} %[0000-0001-8838-1394]
\email{emilie.thelie@austin.utexas.edu}
\affiliation{Department of Astronomy, University of Texas at Austin,
2512 Speedway, Austin, TX 78712, USA}
\affiliation{Cosmic Frontier Center, The University of Texas at Austin, Austin, TX 78712, USA}
\affiliation{Texas Center for Cosmology \& Astroparticle Physics, Austin, TX 78712, USA}

\author{Sarah Libanore}
\affiliation{Department of Physics, Ben-Gurion University of the Negev, Be’er Sheva 84105, Israel}

\author{Yonatan Sklansky}
\affiliation{Department of Astronomy, University of Texas at Austin,
2512 Speedway, Austin, TX 78712, USA}
\affiliation{Cosmic Frontier Center, The University of Texas at Austin, Austin, TX 78712, USA}

\author{Julian B. Muñoz}
\affiliation{Department of Astronomy, University of Texas at Austin,
2512 Speedway, Austin, TX 78712, USA}
\affiliation{Cosmic Frontier Center, The University of Texas at Austin, Austin, TX 78712, USA}
\affiliation{Texas Center for Cosmology \& Astroparticle Physics, Austin, TX 78712, USA}

\author{Ely D. Kovetz}
\affiliation{Department of Physics, Ben-Gurion University of the Negev, Be’er Sheva 84105, Israel}


\begin{abstract}
We propose a new way to reconstruct the ionized-bubble size distribution during the Epoch of Reionization (EoR) through the real-space cross-correlation of 21-cm and star-forming line-intensity maps. Understanding the evolution and timing of the EoR is crucial for both astrophysics and cosmology, and a wealth of information on the first sources can be extracted from the study of ionized bubbles. 
Nevertheless, directly mapping bubbles is challenging due to the high redshifts involved, possible selection biases, and foregrounds in 21-cm maps. 
Here, we exploit the real-space cross-correlation $\xi_{21,\nu}$ between 21-cm and line-intensity mapping (LIM) signals to reconstruct the evolution of bubble sizes during reionization. 
For the first time, we show that $\xi_{21,\nu}(r)$ departs from a saturation level for each separation $r$ when bubbles of size $r$ begin to form, providing a handle for the onset of bubbles of each radius.  
Moreover, we demonstrate that $\xi_{21,\nu}$ evolves from positive to negative as the EoR progresses, reaching a minimum (i.e. maximum anti-correlation) when bubbles of radius $r$ reach peak abundance.
We show that these results are robust to changes in the astrophysical model as well as the timing/topology of reionization. 
This real-space observable complements usual Fourier-space estimators by capturing the localized nature of bubbles, offering new insights into the sources driving cosmic reionization.
\end{abstract}
\vspace*{-.3cm}

%\keywords{dark ages, reionization, first stars - methods:numerical - methods:statistical - cosmology}%Use showkeys class option if keyword
                              
\maketitle

The epoch of reionization (EoR) represents the last major phase-transition of our Universe and a cosmological puzzle to be deciphered \cite{Furlanetto:2006,Wise:2019,Choudhury:2022}. The first stars, galaxies, and black holes emerge during this epoch and ionize their surrounding hydrogen gas, creating ``\textit{ionized bubbles}''. These large-scale ($\sim1-100\,{\rm Mpc}$~\cite{Furlanetto:2004}) HII pockets grow as the EoR proceeds, eventually merging when reionization is complete, around redshift $z\sim 5.5$~\cite{Barkana:2001,Mesinger:2007,Mesinger:2010,McGreer:2015,Davies:2018,Greig:2019}.
While this broad picture holds, many questions remain unanswered: \textit{what was the exact timing of the EoR? Which sources drove its evolution?} 
The growth and distribution of ionized bubbles can unveil the progress of reionization~\cite{MiraldaEscude:2000,McQuinn:2007,Friedrich:2011,Chen2019, Hutter:2020,Gazagnes2021,Giri:2021,Thelie:2023,Giri:2019,Jamieson:2025}, constrain the properties of the first cosmic sources~\cite{Kapahtia:2021,Doussot:2022,Pathak:2022,Thelie:2022,Diao:2024,Schwandt:2025}, and even constrain cosmology~\cite{Thelie:2025}.

The 21-cm signal is a promising probe of ionized bubbles, as it traces neutral hydrogen. 
Multiple radio telescopes  target this observable~\cite{HERA:2023,Mertens:2020,Trott:2020,Parsons:2010,Dilullo:2021}, placing powerful limits on its power spectrum~\cite{HERA:2025} and perhaps providing 21-cm tomographic images in the near future~\cite{Koopmans:2015}.
However, extracting bubbles from 21-cm data is complicated by foregrounds and modeling uncertainties. 
Foregrounds are three-to-four orders of magnitude stronger than the EoR signal~\cite{Bernardi:2009,Bernardi:2010}, requiring careful subtraction or complete avoidance (e.g.\ discarding the contaminated Fourier modes~\cite{Liu:2020}). 
In addition, the 21-cm signal is a complex tracer of the thermal and ionization state of hydrogen, so recovering bubbles requires careful theoretical modeling~\cite{Chen2019,Gorce:2019,Hutter:2020,Gazagnes2021,Giri:2021}, for instance through machine-learning techniques~\cite{Hiegel:2023,Bianco:2021,Bianco:2024,Bianco:2025,GagnonHartman:2021,Kennedy:2024,Sabti:2024jff}.

A compelling way to complement the 21-cm signal is to cross-correlate it with other tracers of reionization.
This includes surveys of high-redshift galaxies~\cite{Hutter:2023,Hutter:2025,Chen:2025,GagnonHartman:2025,MonteroCamacho:2025} and line-intensity mapping (LIM). 
LIM surveys, in particular, capture the integrated redshifted emission of all sources at a specific frequency. They are thus sensitive to the flux emitted by the full galaxy population, including the faintest sources (see reviews in \cite{Kovetz:2017agg,Bernal:2022jap,Chang:2026}).
Several LIM experiments are targeting lines associated with star-formation processes at high $z$, including [OIII], [OII], H$\alpha$, H$\beta$, Ly$\alpha$~\cite{Bock:2025ijl}, CO~\cite{Keating:2020wlx,COMAP:2021nrp} and [CII]~\cite{CONCERTO2020, TIME, Karoumpis_2022}.
Despite the intrinsic challenges that LIM surveys face due to foregrounds and interloper lines~\cite{Breysse:2015baa,Lidz:2016lub,Cheng:2020asz}, they remain a promising probe of the high-$z$ Universe.
A robust body of work on 21-cm $\times$ LIM cross-correlations has shown that they can mitigate systematic effects~\cite{Furlanetto:2006pg,Lidz:2011dx,McBride_2024} and probe the global timing of reionization~\cite{Lidz:2008ry,Moriwaki:2019,Moriwaki:2024kvp,Fronenberg_2024,Kannan:2021ucy,Sun:2024vhy,Chang:2019xgc,Beane_2019,Murmu:2022}.

Recently, we showed in Ref.~\cite{Libanore:2025gte} that the Pearson coefficient (i.e., the zero-lag  real-space cross-correlation) decreases from a saturation value at the onset of EoR and transitions from positive to negative as reionization progresses. That is, LIM and 21-cm maps begin to decorrelate at small scales when the first bubbles appear. 
Building on this result, here we show that real-space 21-cm $\times$ LIM cross-correlations are a powerful tracer of the abundance of ionized bubbles of each size, and thus of the physics of reionization. 
The key insight is that bubbles are real-space objects, and as such the best way to map them is through features in the correlation function, rather than its Fourier-transformed power spectrum. In the following, we show that the cross-correlation at a comoving separation $r$ traces the abundance of bubbles of that approximate radius, probing the emergence of ionized structures and the dominant bubble size at each redshift $z$. 

\paragraph*{\bf Modeling the signal}
The 21-cm signal traces reionization through the neutral-hydrogen fraction $\xHI$, as its brightness temperature is
\begin{equation}\label{eq:T21_def}
    T_{21}(z)= \xHI  \, T_0(1+\delta) \left(1-\frac{\Tcmb}{{T}_S}\right).
\end{equation}
Here $\Tcmb$ and ${T}_S$ are the CMB and spin temperatures, whereas $\delta$ is the baryon overdensity and $T_0 = 23 \text{ mK} \times \sqrt{(1+z)/16}$ a normalization factor (where we fix cosmological parameters to {\it Planck 2018}~\cite{Planck:2020}).  
Galaxies excite, heat, and ionize the surrounding intergalactic medium (IGM), altering both $T_S$ and $\xHI$ while correlating them. 
As such, $T_{21}$ is a non-local and complex tracer of star formation, posing key obstacles to extracting reionization bubbles from this observable alone.
For instance, there can be no $T_{21}$ signal in a region either because $T_S \approx \Tcmb$ or $\xHI \approx 0$.

Fortunately, the same galaxy formation can be traced by other line-intensity maps. The observable here is
\begin{equation}
    \bar{I}_\nu (z) = \frac{c\, \phi_{\rm EtoL}}{4\pi \nu_{\rm{rest}} H(z)} \int d\Mh\frac{dn}{d\Mh}L(\Mh,z),
\end{equation}
where $\nu_{\rm rest}$ is the rest-frame frequency of the target line, $c$ the speed of light, $H$ the Hubble rate, $dn/d\Mh$ the halo mass function and $L(\Mh,z)$ the scaling relation linking the line luminosity to the host dark matter halo mass $\Mh$. The factor $\phi_{\rm EtoL}$ accounts for conversion between Lagrangian and Eulerian space (see~Ref.~\cite{Libanore:2025wtu} for details). 
It is clear that $T_{21}$ and $I_\nu$ carry complementary information.
Regions of high star formation will ionize earlier ($\xHI\to 0$) and also have higher LIM intensity ${I}_\nu$. Combining these observables will thus allow us to extract reionization-bubble statistics.

In order to compute correlation functions we generate realizations of both 21-cm and star-formation tracers (e.g.\ [OIII]) maps. 
For the former we use a soon-to-be-released~\cite{Sklansky:2025} version of $\zeus$ \footnote{\href{https://github.com/JulianBMunoz/Zeus21}{https://github.com/JulianBMunoz/Zeus21}}, that extends the original efficient and analytical model of the Cosmic Dawn \cite{Munoz:2023kkg} into the EoR. The new version of the code provides an end-to-end Cosmic Dawn-to-EoR model to return the 21-cm signal and star-formation rate densities in seconds, allowing fast exploration of the large cosmological and astrophysical parameter spaces.
For the star-formation tracers, we rely on $\olimpus$ \footnote{\href{https://github.com/slibanore/oLIMpus}{https://github.com/slibanore/oLIMpus}}, released in Ref.~\cite{Libanore:2025wtu}. 
This code extends the formalism of $\zeus$ to effectively model the power spectrum of other lines, including [OIII], [OII], H$\alpha$, H$\beta$, CO, [CII] based on a variety of scaling relations calibrated on hydrodynamical simulations and low-redshift data. In the following, we consider $\nu=\text{[OIII]}=4960\,$\AA\, as a case study, in analogy to previous works~\cite{Moriwaki:2019,Libanore:2025gte}; the discussion can be straightforwardly extended to other lines.

\paragraph*{\bf Producing the maps}
We produce (coeval) boxes of 21-cm and [OIII] fluctuations, as well as $\xHII$, which we will use to measure the $T_{21}\times I_{\nu = \rm [OIII]}$ cross-correlation. 
In order to probe bubbles of different sizes, we use two resolutions and box sizes: $(dx,\Lbox) = (1 \text{ cMpc},250  \text{ cMpc})$ and $(2 \text{ cMpc},500  \text{ cMpc})$. 
The former spans bubble radii between $2-10$ cMpc, and the latter between $10-20$ cMpc (and we find good agreement in the overlapping region, more details in Appendix~\ref{app:verification}).
Each set is composed of ten runs, each with a different seed. The 21-cm and [OIII] maps are smoothed with a top-hat kernel of size equal to the box resolution. For now, we will  assume a fiducial astrophysical model where the EoR ends at $z\sim5.8$, and return to vary astrophysical parameters below. Appendix~\ref{app:fiducial} summarizes our astrophysical modeling.

\begin{figure}
\centering
\includegraphics[width=0.4\textwidth]{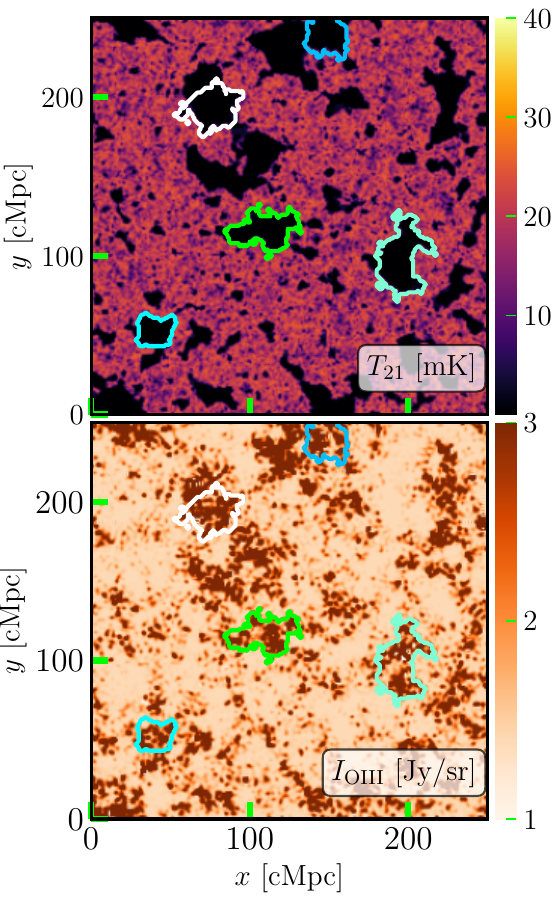}
\caption{Maps of the 21-cm signal $T_{21}$  (\textbf{top}) and [OIII] intensity (\textbf{bottom}) at $z=7.5$ (corresponding to a volume ionized fraction $\xHIIavg=0.29$). The colored contours locate ionized bubbles of effective radius $\Rb \sim 15$ cMpc, and clearly highlight the anti-correlation between the 21-cm and LIM signals at this scale.
\label{fig:maps}
}
\end{figure}

\begin{figure}
\centering
\includegraphics[width=0.5\textwidth]{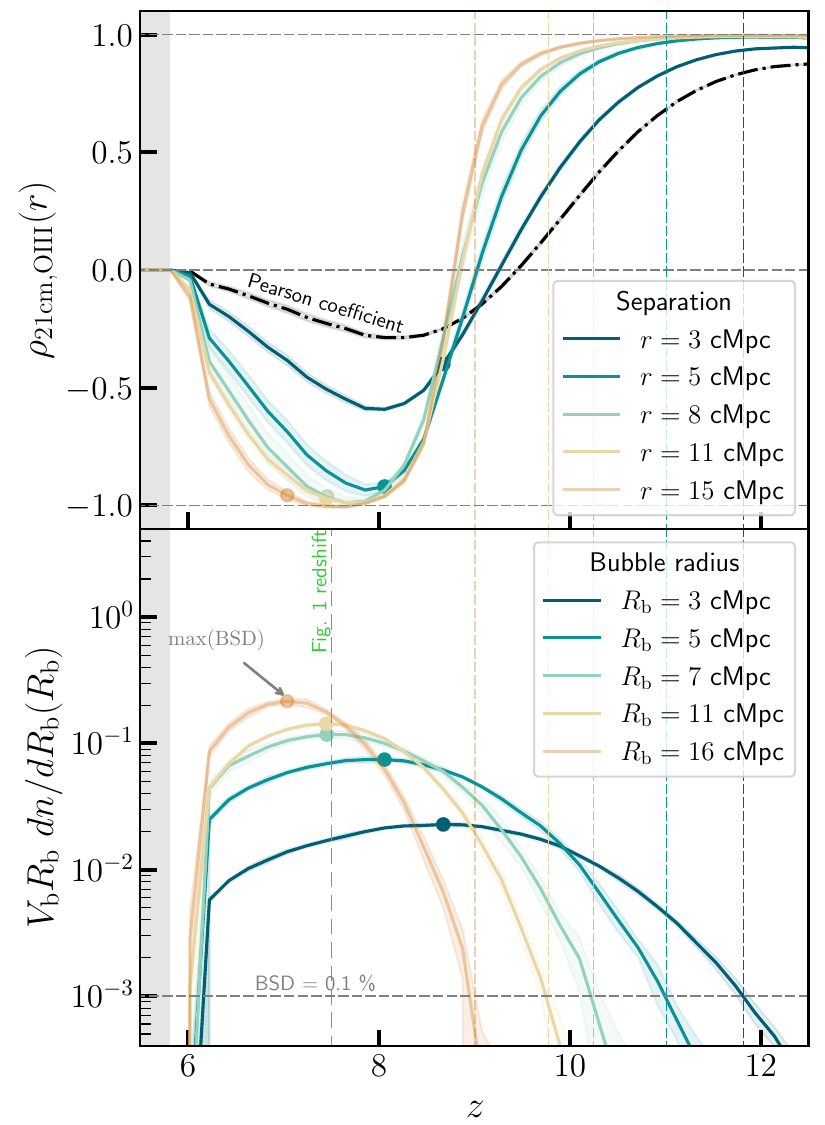}
\caption{Real-space cross-correlation coefficient $\rxi$ (\textbf{top}) at different separations $r$ (in colors) and bubble size distributions  $dn/d\Rb$ (\textbf{bottom}) for different bubble sizes $\Rb$ (in colors, where bubbles need not be spherical, so radius is to be understood as $\Rb\sim\Vb^{1/3}$). 
For reference, the black dashed-dotted line in the top panel shows the Pearson coefficient \cite{Libanore:2025gte}, which is the zero-separation cross-correlation function.
The vertical lines represent the redshift at which 0.1\% of the cosmic volume is occupied by bubbles of each radius (i.e., $\BSD\gtrsim10^{-3}$), which coincide with the point where $\rxi\lesssim 0.9$, illustrating that $\rxi$ can pinpoint the timing at which the first bubbles of a given size appear. Moreover, the colored dots show the peak of the BSD at each size $\Rb$, which are traced by the minimum of $\rxi$ for each $r$, so the cross-correlation is minimal when the BSD peaks. The shaded colored regions represent the scatter between our set of 10 simulations, and reionization is over at $z\sim 5.8$, as indicated by the gray band. 
\label{fig:xi_BSD}
}
\end{figure}

Figure \ref{fig:maps} shows a slice through one of the maps used in our analysis. 
Regions with strong [OIII] emission trace higher star-formation rate density, which ionize first and thus have lower $T_{21}$. To guide the eye, we added colored contours to both maps, delineating bubbles with effective radius $\Rb\sim15$ cMpc (i.e.\ whose volume is that of a sphere with a $\Rb=15$ cMpc radius), corresponding to the typical size of reionization bubbles at this redshift. In these regions, the 21-cm and [OIII] signals are negatively correlated. This anti-correlation has a specific spatial pattern, which we investigate in the following.

\begin{figure}
\centering
\includegraphics[width=0.5\textwidth]{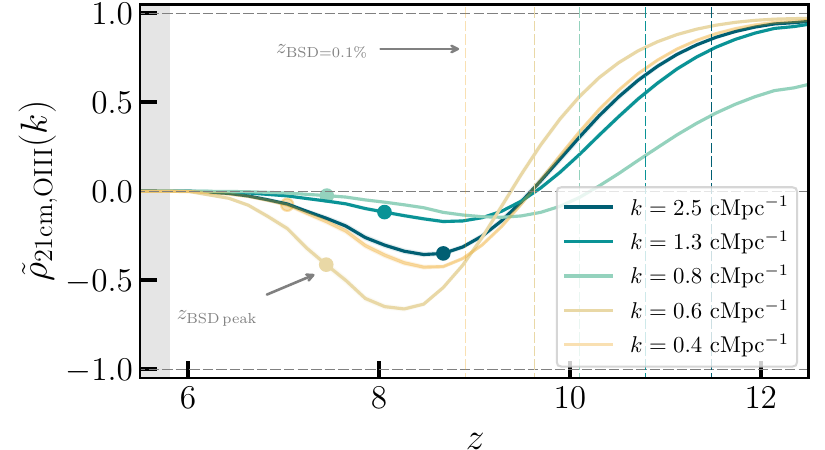}
\caption{Fourier-space cross-correlation coefficient $\rdelta$ at different wavenumbers $k=2\pi/r$ (in colors). 
As in Fig.~\ref{fig:xi_BSD}, the vertical lines represent the redshift at which $\BSD(\Rb)=10^{-3}$ and the colored dots correspond to the redshifts at which the BSD peaks. This figure shows how the sharp decline of correlations at different separations seen in Fig.~\ref{fig:xi_BSD} is erased in Fourier space, due to mode mixing.
\label{fig:Delta2}
}
\end{figure}

\paragraph*{\bf Scale-dependent cross-correlation}
Our main observable is the real-space cross-correlation coefficient, defined as
\begin{equation}\label{eq:rho21oiii}
    \rxi (r,z) = \frac{\xicross(r,z)}{\sqrt{\xi_{\rm{21cm}}(r,z) \xi_{\nu}(r,z)}},
\end{equation}
which depends on the real-space two-point cross-correlation function between the 21-cm and star-forming lines denoted by their rest-frame frequency $\nu$:
\begin{equation}
    \xicross (r,z) = \langle \delta_{\rm{21cm}}(x,z) \delta_{\nu}(x+r,z) \rangle,
\end{equation}
where $\delta_{\rm{21cm}}$ and $\delta_{\nu}$ are the 21-cm and star-forming line (here [OIII]) fluctuations. $\xi_{\rm{21cm}}$ and $\xi_{\nu}$ are the real-space auto-correlation functions of the 21-cm and $\nu$ lines respectively, and $r$ is the separation scale.

We show the cross-correlation coefficient computed from our fiducial set of simulations in the top panel of Fig. \ref{fig:xi_BSD} as a function of redshift $z$ for several separations $r$. Globally, the cross-correlation coefficient shows a similar behavior against $z$ at all separations $r$. 
Before the onset of reionization ($z\gtrsim12$), 21-cm and [OIII] signals are highly correlated, so $\rxi\approx 1$. This is because overdense regions have more [OIII] emission as well as a more positive $T_{21}$ (due to X-ray heating). 
Once reionization begins, however, the cross-correlation coefficient drops, as ionized bubbles (which emit no 21-cm signal, but are still star-forming) begin to appear. This marks the onset of reionization \citep{Libanore:2025gte}.
As the universe becomes more ionized, the cross-correlation first crosses zero and then turns negative \citep{Moriwaki:2019}. Around $z\sim6-8.5$ (corresponding to ionized fractions $\xHIIavg\sim0.1-0.4$), the bubbles fill most of the star-forming regions (so $T_{21} = 0$ at the location of the peaks in the [OIII] map), creating an anti-correlation that deepens as the bubbles grow.
The bubbles ($T_{21}=0$ regions) then fill more and more space, which reduces the anti-correlation, so that at the end of the EoR there is no correlation at all ($\rxi\to 0)$.
In Appendix \ref{app:verification}, we show the complementary view with $\rxi$ against separation $r$. We note that Ref.~\cite{Hutter:2023} performed a similar analysis by cross-correlating the 21-cm signal against Lyman-$\alpha$ emitting galaxies.

The spatial behavior of the cross-correlation coefficient $\rxi$ holds additional information. In the top panel of Fig.~\ref{fig:xi_BSD}, the lines with different colors represent different separations $r$ (along with the Pearson coefficient, which corresponds to zero separation). 
We can see that the 21-cm and [OIII] signals decorrelate ($\rxi$ deviates from unity) first for smaller separations $r$. 
That is because small reionization bubbles appear before larger ones. 
Moreover, $\rxi(r)$ reaches its minimum at different redshifts for different separations $r$. 
The small scales anti-correlate earlier, but do not reach $\rxi= -1$. The timing has the same explanation as above, but the higher minimum suggests that the density structures (with the larger densities being also more star-forming) that correlate the 21-cm and [OIII] signals prevent $\rxi = - 1$ for the smaller scales.

From this first global analysis of the cross-correlation coefficient, we can already see that it contains valuable information about the onset and abundance of ionized bubbles. 
In order to quantify this insight, we extract bubbles from our simulations using the watershed algorithm \cite{Lin:2016}, and compute the bubble size distributions (BSDs) at each redshift. 
We note that watershed is not the only way to obtain the bubbles sizes (see e.g.\ \cite{Friedrich:2011,Lin:2016,Giri:2019,Bianco:2021}), but we leave the exploration of other ways of measuring BSDs for future work.
The bottom panel of Fig.~\ref{fig:xi_BSD} shows $\BSD$, with $dn/d\Rb$ the BSDs and $\Rb$  and $\Vb$ the radius and volume of bubbles, respectively ($\Rb\sim\Vb^{1/3}$). This quantity is dimensionless and can be understood as the fraction of the volume of the universe that is occupied by ionized bubbles of size $\Rb$.
As expected, the smallest bubbles appear first and grow as reionization progresses.

Comparing the top (cross-correlation coefficient) and bottom (bubble size distributions) panels of Fig. \ref{fig:xi_BSD}, we identify two important features:
\begin{enumerate}
    \item \textit{Emergence of ionized bubbles:} The cross-correlation coefficient $\rxi(r,z)$ departs from unity at the redshift $z_{\rxi=0.9}$ at which the first bubbles of size $r=\Rb$ appear. Quantitatively, we find that at redshift $z_{\rm BSD=0.1\%}$ when
    bubbles of radius $\Rb$ cover roughly 0.1\% of the cosmic volume ($\BSD(\Rb,z)=1\times10^{-3}$) the cross-correlation coefficient drops to $\rxi(r,z)\sim0.9$. 
    
    \item \textit{Bubble sizes:} The cross-correlation coefficient $\rxi(r,z)$ reaches its minimum when the bubbles of size $r\sim\Rb$ are most abundant. 
    That is, the redshift $z_{\rm BSD\,peak}$ at which $\BSD(\Rb)$ peaks corresponds to the redshift  $z_{\rm corr\,min}$, where  $\rxi(r=R_b)$ is at its minimum. 
\end{enumerate}
These two features provide a new avenue for characterizing the EoR: the real-space cross-correlation anchors the abundance of bubbles of a certain size (for $\Rb\lesssim 15$ cMpc, as for larger bubbles the estimators lose precision).   
For instance, by comparing when the $r=3$ and 10 cMpc cross-correlation coefficient  drops to $\rxi(r,z)=0.9$, we can test how delayed is the onset of large bubbles compared to smaller ones, and thus infer the shape and growth of the BSD.
Reconstructing the BSD, in turn, will allow us to infer the properties of the underlying astrophysical sources, as we will discuss below when we vary the astrophysical modeling parameters.
The accuracy of such reconstruction in our model can be estimated from the differences $\delta z_1 = z_{\rxi=0.9}-z_{\rm BSD=0.1\%}$ and $\delta z_2 = z_{\rm corr\,min}-z_{\rm BSD\,peak}$; in the case of Fig.~\ref{fig:xi_BSD}, these correspond to $\delta z_1\lesssim10 \%$ and $\delta z_2 \lesssim10 \%$. We discuss the uncertainties of these two redshift estimators in Appendix \ref{app:redshiftmatch}, where we show $z_{\rxi=0.9}$ versus $z_{\rm BSD=0.1\%}$ and $z_{\rm corr\,min}$ versus $z_{\rm BSD\,peak}$.

Clearly, we can extract bubble information from the real-space cross-correlation coefficient. 
Before discussing the implications of this result, let us compare to past studies in Fourier space and determine whether our result is robust in other astrophysical scenarios.

\paragraph*{\bf Fourier-space cross-correlation} Previous studies (e.g.\ \cite{Lidz:2008ry,Moriwaki:2019})  analyzed the Fourier-space cross-correlation coefficient $\rdelta(k)$ (see Appendix~\ref{app:verification} for its definition and further discussion), finding that its transition from positive to negative  traces the mid-point of reionization, when the ionized volume fraction is $\bar{x}_{\rm HII}^{\rm V}\sim 30\%-50\%$.

The natural question, therefore, is whether the Fourier-space coefficient $\rdelta(k)$ can also be used to trace the evolution of the BSD. To answer this, in Fig.~\ref{fig:Delta2} we show $\rdelta(k)$ at different scales $k$, chosen to match $2\pi/r$ for the values of $r$ shown above. 
This Fourier-space estimator manifestly loses the bubble information that the real-space estimator was featuring: at the beginning of the EoR, all different $\rdelta(k)$ lines depart from 1 approximately at the same redshift. Moreover, the redshift position of the minimum of $\rdelta(k)$ no longer corresponds to the redshift position of the BSD peaks (marked by the dots). We conclude that the mixing of real-space scales that occurs in the Fourier transform (see Eq.~\ref{eq:fourier}) erases the two features imprinted by bubbles, making the Fourier-space cross-correlation less efficient than the real-space one in the reconstruction of real-space objects.

\paragraph*{\bf Varying astrophysical parameters}
\label{sec:varying_astro}

Thus far we have shown results for a single fiducial reionization model, detailed in Appendix~\ref{app:fiducial}. Now, we briefly explore whether varying key astrophysical quantities impacts our results.

One of the main unknowns of EoR modeling is the escape fraction $\fesc$ of ionizing photons from the first galaxies into the IGM. In $\zeus$, it is assumed to behave as
\begin{equation}
    \fesc (\Mh) = \fesczero \left( \Mh/10^{10}\Msun \right)^{\alphaesc},
\end{equation}
where $\fesczero$ is a normalization parameter and $\alphaesc$ the power-law index against halo mass. 
Our fiducial case has $\fesczero=10\%$ and $\alphaesc=0$, meaning that all galaxies have the same escape fraction. Positive (negative) $\alphaesc$ leads to massive galaxies having more (less) ionizing photons escaping, with a consequent impact on the bubble evolution. 
Another uncertainty arises from the X-ray luminosity per unit SFR $\LX$ (defined in $\log_{10}$ erg/s/($\Msun$/yr)). Low X-ray luminosities have been disfavored by HERA results~\cite{HERA:2023} (see also \cite{Lazare:2023jkg}), hence we set our fiducial value to $\LX=41$.

These are the main parameters that define the sources of reionization in our model, and thus its timing and topology. 
We vary one parameter at a time, and thus generate thirty more simulations with $\zeus+\olimpus$: one for each box size, each varying $\fesczero = [0.01,0.5]$, $\alphaesc =[-0.5,0.5]$, and $\LX = [39.5,42.5]$.

\begin{figure}
\centering
\includegraphics[width=0.49\textwidth]{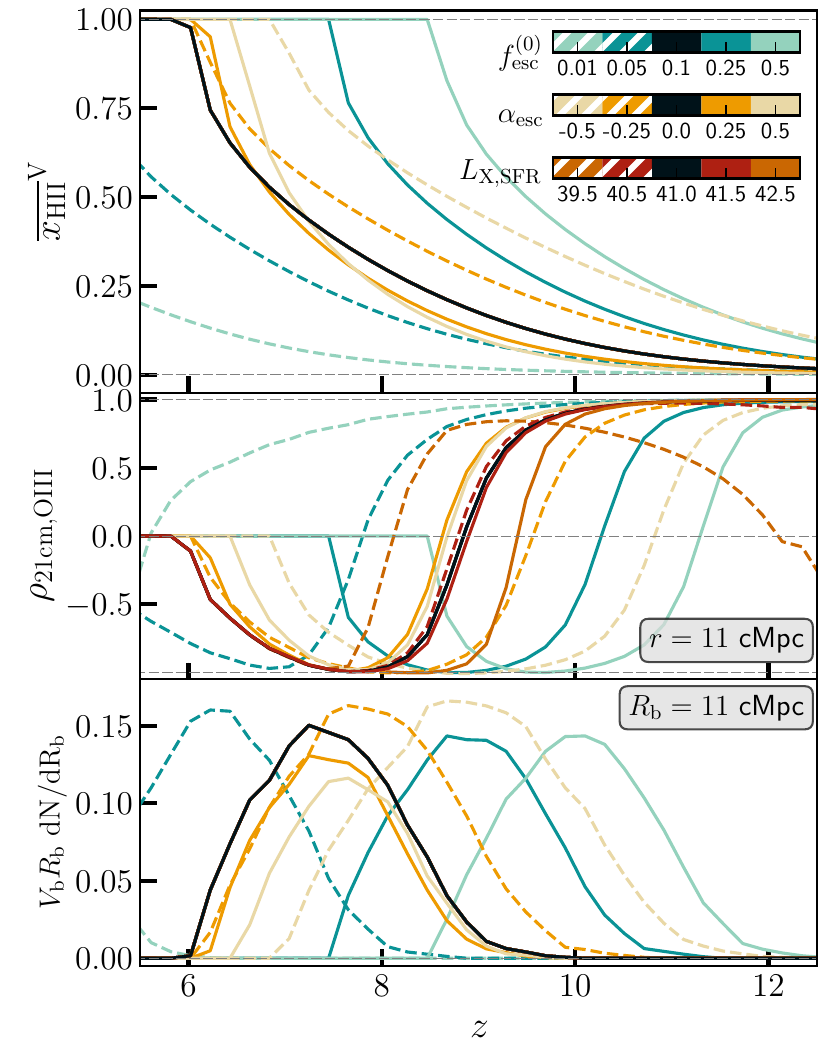}
\vspace{-0.25in}
\caption{We test that our results hold when varying the timing and topology of reionization through changes in the escape fraction amplitude $\fesczero$ (in blue), its power law index $\alphaesc$ (in yellow), and the X-ray luminosity $\LX$ (in red). 
We show the volume fraction of ionized hydrogen $\xHIIavg$ (\textbf{top}), real-space cross-correlation coefficient $\rxi$ (\textbf{middle}), and BSDs (\textbf{bottom}). In all cases (except disfavored low $\LX\lesssim 39.5$ \cite{HERA:2023}, red-dashed) we find a strong correlation between the departure of $\rxi(r=11$cMpc) from unity (middle panel) and the onset of the first 11-cMpc bubbles (bottom). Likewise, the BSD peaks where $\rxi$ reaches its minimum. 
\label{fig:varying_astro_xHII_r_BSD}
}
\vspace{-0.25in}
\end{figure}

The results of this test are summarized in Fig.~\ref{fig:varying_astro_xHII_r_BSD}.
The top panel shows the volume-averaged ionized fraction, $\bar{x}_{\rm HII}^{\rm V}$ in each model. 
As expected, $\fesczero$ shifts the timing of reionization, with small (large) values leading to a later (earlier) EoR. Negative $\alphaesc$ induce a more gradual reionization driven by fainter galaxies, while positive $\alphaesc$ accelerate it. On the other hand, the X-ray luminosity $\LX$ has negligible impact on the ionized field, and thus on $\xHIIavg$; however, its value affects the evolution of $T_{21}$, and hence the shape of the cross-correlation coefficient. 

We show this coefficient in the central panel of Fig.~\ref{fig:varying_astro_xHII_r_BSD} for $r=11$\,cMpc and for each parameter set. The coefficient is shifted in redshifts when the escape fraction parameters are varied, and shows little difference when varying $\LX$; in fact, at these redshifts the evolution of $\rho_{\rm 21cm,OIII}$ is dominated by bubble formation, while the role of $T_{21}$ is crucial at higher $z$, before $\rho_{\rm 21cm,OIII}=1$ is reached. The only exception is found for the smallest value of $\LX=39.5$; as already discussed in Ref.~\cite{Libanore:2025gte}, this corresponds to a scenario in which EoR begins in a still-cold IGM. This scenario affects the effectiveness of our real-space estimators, since it prevents the cross-correlation from  reaching  saturation before the bubble begins to form. Fortunately, this scenario is disfavored \cite{HERA:2023} and therefore does not affect our conclusions. 

Finally, we extract bubbles using the watershed algorithm, and show the BSD evolution (at $\Rb\!=\!11$ cMpc) for all the models (bottom panel of Fig.~\ref{fig:varying_astro_xHII_r_BSD}).
In all cases, we see an excellent match between the onset of the $\Rb\!=\!11$ cMpc bubbles and the departure from $\rxi\!=\!1$ at the same separation $r\!=\!\Rb$.
Moreover, the cross-correlation coefficient is minimized when the BSD is maximized. 
We conclude that the impact on the reionization process of changes in the astrophysical parameters is reflected in the cross-correlation coefficient as well, which thus conserves the information on bubbles for every model tested here (with $\LX=39.5$ the only exception).

\paragraph*{\bf Discussion \& Conclusion}
\label{sec:conclusion}

In this Letter, we have shown that the real-space cross-correlation coefficient $\rxi(r,z)$ 
between 21-cm and star-forming line-intensity maps directly traces ionized bubbles during the EoR.
Because ionized bubbles are localized structures, this coefficient provides a more direct and efficient probe than Fourier-space observables. 
While this work serves as a proof-of-concept, rather than a full pipeline to extract the BSD at high precision, we have tested that our results are robust to variations in reionization timing and topology across our astrophysical parameter space. 
Our conclusions agree with previous studies of cross correlations of 21-cm with [OIII] and  Lyman-$\alpha$ emitters~\cite{Moriwaki:2019,Hutter:2023}.

Our main finding is that $\rxi(r,z)$ drops from unity and reaches $\sim 0.9$ when the first bubbles of radii $\Rb=r$ appear. This is due to ionized bubbles emitting no 21-cm signal but still being dense, star-forming regions. 
Moreover, the cross-correlation coefficient is minimized when the BSD of bubbles of $\Rb=r$ reaches its peak. Again, bubbles are regions of high star formation, so when bubbles of size $r$ are the most abundant, they will induce a strong anti-correlation at the same $r$ separation. 
These features of the cross-correlation coefficient provide a new way to measure the BSD and thus constrain the topology and sources of reionization.

In future work we will estimate the sensitivity of real-space cross-correlations to the sources of reionization and their clustering, with forecasts for specific experiments. 
We strongly encourage further studies of these cross-correlations, including modeling the impact of foregrounds and instrumental noise, extending beyond two-point functions, and building on the current versions of $\zeus$ and $\olimpus$ to analytically compute the cross-correlation coefficient.

We conclude that real-space cross-correlations provide a powerful tool to understand the drivers of reionization through the bubbles they create. With a wealth of 21-cm and star-forming LIM observations on the horizon, this observable offers a direct window into bubble sizes and evolution that complements Fourier-space analyses.

\begin{acknowledgments}
We thank Joohyun Lee and Paul Shapiro for discussions. 
This work has been supported at UT Austin through NSF Grants AST-2307354 and AST-2408637, and the CosmicAI institute AST-2421782. 
This research was supported in part by grant NSF PHY-2309135 to the Kavli Institute for Theoretical Physics (KITP).
SL thanks the Azrieli Foundation for support through an Azrieli International Postdoctoral Fellowship.
EDK acknowledges
 support from the U.S.-Israel Bi-national Science
Foundation (NSF-BSF grant 2022743 and BSF grant 2024193) and the Israel National Science Foundation (ISF grant 3135/25), as
well as support from the joint Israel-China   program (ISF-NSFC grant  3156/23). 
\end{acknowledgments}

\bibliography{biblio}% Produces the bibliography via BibTeX.

\appendix

\section{Consistency Checks}
\label{app:verification}

This appendix presents consistency checks and additional results obtained from our simulations.

\begin{figure}[h!]
\centering
\includegraphics[width=0.4\textwidth]{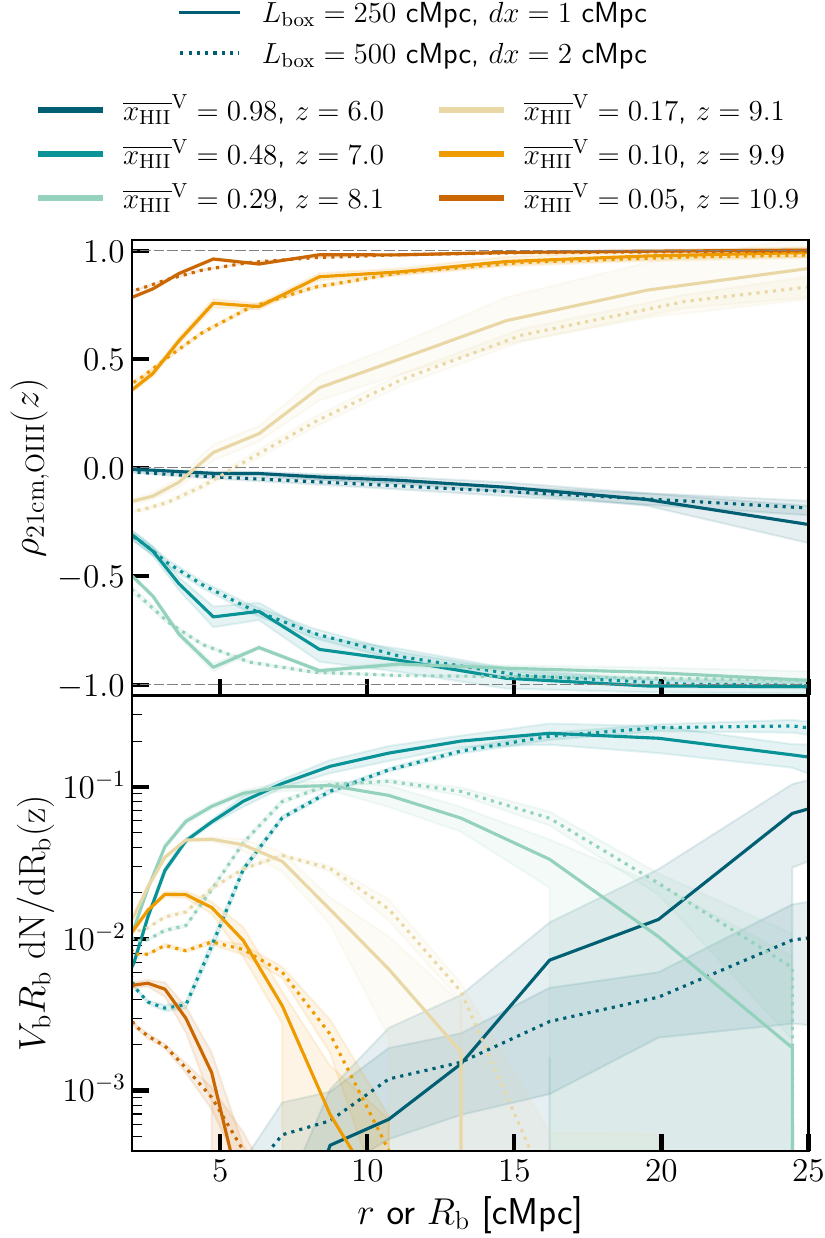}
\caption{Real-space cross-correlation coefficient $\rxi$ (\textbf{top}, vs separation $r$) and bubble size distributions (\textbf{bottom}, vs size $R_b$), both shown at different redshifts $z$ (in colors). The solid and dotted lines show the two sets of simulations, the first one with $(dx,\Lbox) = (1 \text{ cMpc},250  \text{ cMpc})$ and the second one with $(dx,\Lbox) = (2 \text{ cMpc},500  \text{ cMpc})$. 
\label{fig:xi_BSD_vs_scale}
}
\end{figure}

\begin{figure}[h!]
\centering
\includegraphics[width=0.4\textwidth]{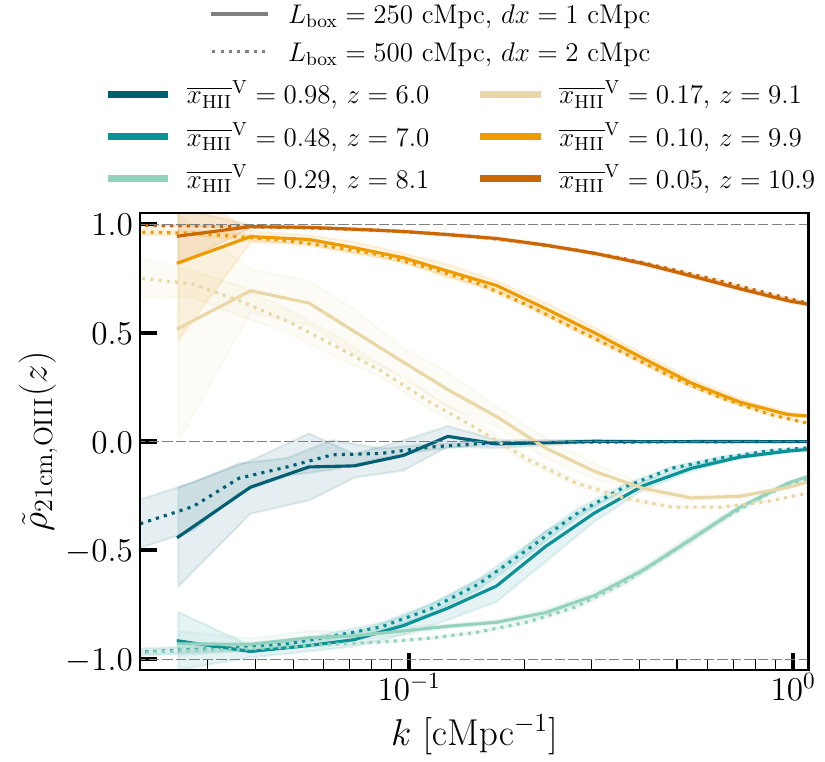}
\caption{Fourier-space cross-correlation coefficient $\tilde \rho_{\rm 21cm,OIII}$ at different redshifts $z$ (in colors). The full and dotted lines show the two sets of simulations as in Fig.~\ref{fig:xi_BSD_vs_scale}.
\label{fig:Delta2_vs_scale}
}
\end{figure}

\begin{itemize}
    \item[]{\bf Pearson coefficient --} Our analysis extends the results of Ref.~\cite{Libanore:2025gte}. As shown in Fig.~\ref{fig:xi_BSD}, the Pearson correlation coefficient corresponds to the $r\to 0$ limit of the real-space cross-correlation function. Its amplitude (similarly to $\rho_{\rm 21cm,OIII}$ at small $r$) is smaller than unity due to decorrelation induced by shot noise. As $r$ increases, the effect of shot noise is progressively smoothed out, and the correlation approaches unity. Because it encodes local information, the Pearson coefficient departs from this saturation earlier and reaches negative values at smaller separations than the real-space cross-correlation.

    \item[]{\bf Cross-correlation coefficient and BSDs with respect to scale --} For completeness, Fig.~\ref{fig:xi_BSD_vs_scale} presents $\rho_{\rm 21cm,OIII}$ as a function of $r$ instead of $z$. This configuration helps to visualize the transition from positive to negative correlation; for small bubbles, this  takes place at the beginning of the EoR, when $\overline{x_{\rm HIII}}^{\rm V}\sim 10\%$ ($z\sim 8.7$), while it happens later for larger scales. 
    The plot also shows the bubble size distribution (bottom panel), and  highlights its evolution. Initially, the distribution is dominated by small bubbles, while large bubbles become more and more dominant at lower $z$. Also, $\rxi(z)$ flips sign at any scales $r$ when the average IGM ionizes significantly since the presence of bubbles of size $r$ anti-correlate the signals.
    Finally, the plot compares results from the two sets of coeval boxes adopted in the main analysis. The larger, lower resolution box underestimates the abundance of small bubbles. On the other hand, the smaller, higher resolution box (1 cMpc, 250 cMpc) overestimates the abundance of large bubbles at high $z$, and it underestimates it at low $z$. On intermediate scales, $r\sim 10\,$Mpc, the two sets are in agreement accounting for variance in the simulations. Thus, in our main analysis, we combine their outputs.
    
    \item[] {\bf Cross-correlation power spectrum --} Past literature have argued that anti-correlation in the power spectrum can be used to track the evolution of the EoR. The estimator advocated  in Ref.~\cite{Moriwaki:2019}, for instance,is the Fourier-space analogous of our Eq.~\eqref{eq:rho21oiii}, namely 
    \begin{equation}
    \rdelta (k,z) = \frac{\Deltacross(k,z)}{\sqrt{\Delta^2_{\rm{21cm}}(k,z) \Delta^2_{\nu}(k,z)}}, 
\end{equation}
which is defined in terms of the two-point auto- and cross-power spectra:
\begin{equation}\label{eq:fourier}
\begin{aligned}
    \Delta_{ij}^2 (k,z) &= \frac{k^3}{2\pi^2} P(k,z) = \langle\delta_{i}^*(k)\delta_{j}^*(k')\rangle\\
    &= \frac{2k^3}{\pi} \int \xi_{ij}(r,z) \frac{\sin{(kr)}}{kr} r^2dr,
\end{aligned}
\end{equation}
with $k=2\pi/r$ the wavenumber, and $\delta_{i,j}^*$ the Fourier transform of the $i,j=$21-cm or LIM signal over-densities respectively. Fig.~\ref{fig:Delta2_vs_scale} shows $\tilde{\rho}_{\rm 21cm,OIII}$ as function of $k$ for different redshifts (and can be directly compared with Fig. 3 in Ref.~\cite{Moriwaki:2019} for consistency). This Figure shows positive correlation at all $k$s for $z\geq 10$, which quickly transitions to negative correlation on small scales (high $k$s) as the EoR proceeds. These scales are also the first where the signal nullifies. The plot shows that the Fourier-space cross-correlation coefficient carries  information akin to our real-space estimator. However, as discussed in the main text, the relation between $k$s where $\tilde{\rho}_{\rm 21cm,OIII}(k,z)=0$ and the bubble size distribution is less straightforward.
Of course, information is preserved when transforming from real to Fourier space, but  mode mixing in Eq.~\eqref{eq:fourier} makes Fourier-space a less direct probe of localized bubbles.  
\end{itemize}

\section{Fiducial Astrophysical Model}
\label{app:fiducial}

The coeval boxes analyzed in the main text have been produced relying on the fiducial astrophysical model implemented in \zeus\, and \olimpus. Table~\ref{tab:fiducial_pars} summarizes the main parameters required to compute the 21-cm and [OIII] signals, and the fiducial values we adopted throughout our work. The star-formation rate is based on Ref.~\cite{Sabti:2021xvh}, while the line intensity on Ref.~\cite{yang2025newframeworkismemission}. The interested reader can find all the details in Refs.~\cite{Munoz:2023kkg,Libanore:2025wtu}.

\begin{table}[h!]
    \centering
    \begin{tabular}{|c|cc|}
    \hline
        Parameter & Fiducial Value & Scope \\
    \hline
        \multirow{2}{*}{$\epsilon_*$} & \multirow{2}{*}{0.1}& Star-formation rate efficiency\\
        & & for halos with $M_h=10^{10}M_\odot$\\
        \multirow{2}{*}{$\alpha_*$} & \multirow{2}{*}{0.5} & Star-formation rate power-law \\
        & & index on the low-$M_h$ end\\
        \multirow{2}{*}{$\beta_*$} & \multirow{2}{*}{-0.5} & Star-formation rate power-law \\
        & & index on the high-$M_h$ end\\
        \hline
        \multirow{2}{*}{$f_{\rm esc}^{(0)}$ (*)} & \multirow{2}{*}{0.1} & Ionizing photons escape fraction \\
        & & from halos with $M_h=10^{10}M_\odot$ \\
        $\alpha_{\rm esc}$ (*) & 0 & Escape fraction power-law index \\
        \hline
        \multirow{3}{*}{$L_{\rm X,SFR}$ (*)} & \multirow{3}{*}{41} & X-ray luminosity at 2\,keV \\
        & & per unit star-formation rate, \\
        & & defined in $\log_{10}$ erg/s/($\Msun$/yr)\\
    \hline
    \end{tabular}
    \caption{Fiducial values of the astrophysical parameters adopted in the baseline simulation in the main text. Parameters indicated with (*) have been varied to test the stability of our conclusions to variations in the underlying astrophysics.}
    \label{tab:fiducial_pars}
\end{table}

\begin{figure}[hb!]
\centering
\includegraphics[width=0.35\textwidth]{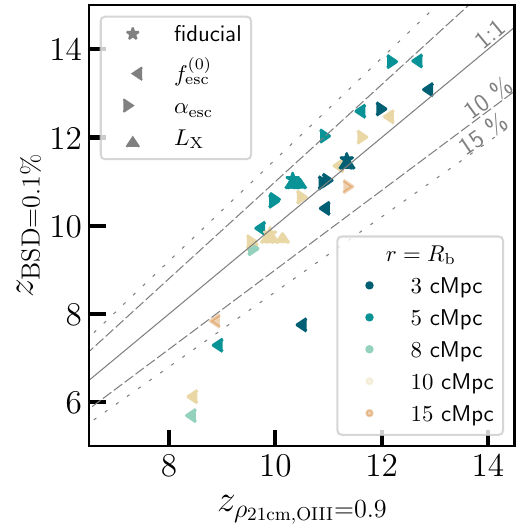}
\caption{Redshift $z_{\rm BSD=0.1\%}$ at which the BSDs are equal to 0.1 (i.e.\ bubbles fill 0.1 \% of the total volume) with respect to the redshift $z_{\rxi=0.9}$ marking the departure of $\rxi$ from unity to the value 0.9. These are shown for different scales (in colors), as well as for all reionization models.
}
\label{fig:z_corr09_bsd01}
\end{figure}

\section{Redshift Estimator}\label{app:redshiftmatch}

\begin{figure}[b!]
\centering
\includegraphics[width=0.375\textwidth]{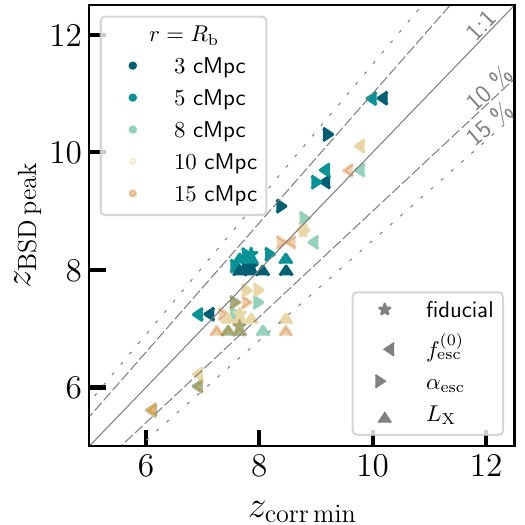}
\caption{Redshift $z_{\rm BSD\, peak}$ at which the BSDs peak (bubbles being the most abundant) with respect to the redshift $z_{\rm corr\,min}$ at which $\rxi$ is minimized (maximum anti-correlation). These are shown for different scales (in colors), as well as for all reionization models.
}
\label{fig:z_corrmin_bsdpeak}
\end{figure}

In the main text, we discussed how to use the redshift $z_{\rho_{\rm 21cm,\nu}(r)=0.9}$ where the cross-correlation coefficient departs from saturation at each scale $r$ to estimate when bubbles with radius $\Rb=r$ begin to form (i.e. the redshift $z_{\rm BSD=0.1\%}$ at which they take 0.1 \% of the cosmic volume). Similarly, we discussed how the redshift $z_{\rm corr\,min}$ of the minimum of the cross-correlation tracks the redshift $z_{\rm BSD\,peak}$ where the BSD peaks at a scale $r$.
Here we show  that these features hold for different scales $r\lesssim 15$ cMpc, and astrophysical scenarios. Figs.~\ref{fig:z_corr09_bsd01} and~\ref{fig:z_corrmin_bsdpeak} summarize our tests. 
Fig.~\ref{fig:z_corr09_bsd01} focuses on the bubble onset: a majority of models show an agreement of $\lesssim 10\%$ between the observable ($z_{\rho_{\rm 21cm,\nu}=0.9}$) and the reconstructed ($z_{\rm BSD=0.1\%}$) redshifts.  
The match in even more stringent between $z_{\rm corr\,min}$ and $z_{\rm BSD\, peak}$, where each model leads to a $\leq 15\%$ difference, as shown in Fig. \ref{fig:z_corrmin_bsdpeak}. 
Together, these Figures show how constraining features on $\rho_{\rm 21cm,\nu}$ will indeed be informative about the bubbles appearance and size.  
We leave for future work an optimized version of these estimators that best captures any $R_b$ dependence.

\end{document}